\documentclass[aps,prl,reprint,showpacs,groupedaddress]{revtex4-1}

\usepackage{amsmath}
\usepackage{amssymb}
\usepackage{hyperref}
\usepackage{graphicx}
\usepackage[capitalize]{cleveref}
\hypersetup{
    colorlinks,
    citecolor=blue,
    linkcolor=blue,     urlcolor=blue
}

\newcommand{\Lsat}[0]{L_\mathrm{sat}}
\newcommand{\Nsat}[0]{N_\mathrm{sat}}
\newcommand{\dkb}[0]{\partial_\xi k_{\beta}}

\begin{document}


\title{Saturation of the hosing instability in quasi-linear plasma accelerators}


\author{R. Lehe}
\email[]{rlehe@lbl.gov}
\author{C. B. Schroeder}
\author{J.-L. Vay}
\author{E. Esarey}
\author{W. P. Leemans}
\affiliation{Lawrence Berkeley National Laboratory, Berkeley, CA
  94720, USA}


\date{\today}

\begin{abstract}
The beam hosing instability is analyzed theoretically for a witness beam in the
quasi-linear regime of plasma accelerators. In this regime, the hosing instability
saturates, even for a monoenergetic bunch, at a level much less than standard
scalings predict. Analytic expressions are derived for the saturation distance
and amplitude and are in agreement with numerical results. Saturation is due to
the natural head-to-tail variations in the focusing force, including the
self-consistent transverse beam loading.
\end{abstract}



\maketitle

The beam hosing instability is a major concern for both conventional accelerators
and plasma-based accelerators. In both cases, this instability can exponentially
amplify the small misalignments between the beam and the accelerating structure,
and potentially lead to a strong degradation of the beam emittance, or even to
complete disruption of the beam. Thus the hosing instability (which is related
to the well-known beam-breakup instability) has been actively
studied in its various regimes \cite{LauPRL1989}.
This includes the long-bunch, weakly-coupled regime \cite{NeilPartAccel1979,WhittumJPhysA1997}
applicable to low-current bunches in conventional accelerators, as well as the long-bunch
\cite{PanofskyRSI1968,WhittumPRL1991,SchroederPRE2012} and short-bunch
\cite{ChaoNIM1980,GeraciPoP2000,DoddPRL2001,SchroederPRL1999,MehrlingPRL2017,HuangPRL2007}
strongly-coupled regimes, which are of interest for beams propagating in plasmas
or ion channels.

In particular, the short-bunch, strongly-coupled regime is relevant to
the evolution of the witness beam in plasma-wakefield acceleration
(both for the beam-driven and laser-driven scheme). In this case, the
exponential growth of the hosing instability as a function of the
acceleration distance has raised concerns regarding the feasibility of
a plasma-based particle collider
\cite{LebedevReview2016}.
However, these predictions have been made in the context of the blow-out
(or bubble) wakefield regime, \textit{i.e.} when the driver
(beam or laser) is strong enough to expell all the plasma electrons, forming
a co-moving ion cavity
\cite{MoraPoP1997,RosenweigPRA1991,PukhovAPB2002,LuPRL2006,TzoufrasPRL2008}.
Plasma accelerators may also operate in the quasi-linear regime, where the
driver excites a plasma density perturbation that is a fraction of
the background plasma density \cite{EsareyRMP2009}.
In this Letter, we show that the sustained exponential growth of the
hosing instability, which indeed applies for monoenergetic beams in the blow-out regime,
does not occur in the quasi-linear regime. Instead, in the
quasi-linear regime, the instability can rapidly saturate, and leads only
to a moderate amplification of the beam misalignment.

This early saturation is due to the head-to-tail spread in betatron
frequency that naturally occurs across the bunch in the quasi-linear
regime. It is indeed well-known that a head-to-tail spread in
betatron frequency can mitigate the hosing instability \cite{LauPRL1989,BalakinProc1983}.
However, in the blow-out regime, the focusing force of the wakefield
is independent of the longitudinal coordinate, and thus any head-to-tail spread
in betatron frequency necessarily requires an energy spread in the
bunch. Owing to the typically large beam-loading for high efficiency, large
energy spreads (e.g. a few percents \cite{MehrlingPRL2017}) are required
\cite{LebedevReview2016}. This is impractical, since many applications of plasma-wakefield
accelerators require monoenergetic beams. By contrast, in the quasi-linear regime,
the focusing force naturally varies as a function of the longitudinal
coordinate, and moreover this variation can potentially be tailored by
beam loading \cite{KatsouleasPartAcc1987}. Therefore, in the
quasi-linear regime, no energy spread is required in order to mitigate
the hosing instability.

\emph{Hosing equation -} In order to study the hosing instability, let us consider the
equation of evolution for the centroid of the witness beam.
For simplicity, the effects of beam acceleration are not considered here, but
can be obtained by a simple change of variable \cite{CherninPartAcc1989,SchroederPRL1999}
[i.e. by replacing, in Eq.~(\ref{eq:evolution}) below, $x_c$
by $\tilde{x}_c = (\gamma(z)/\gamma(0))^{1/4}x_c$ and
$z$ by $\tilde{z} = \int_0^z (\gamma(0)/\gamma(z'))^{1/2} dz'$].
Under these assumptions, the
transverse evolution of the centroid $x_c(\xi, z)$ is governed by the equation
\begin{eqnarray}
\label{eq:evolution}
\partial_z^2 &&x_c(\xi, z)+ k_\beta^2(\xi)\,x_c(\xi, z) = \nonumber \\
&& k_c^2\int_{\xi}^0\! \frac{n_b(\xi')}{n_p} x_c(\xi',z) \sin[\kappa_p(\xi'-\xi)] \,\kappa_p d\xi',
\end{eqnarray}
where $z$ is the propagation distance, and $\xi = z-ct$ is the head-to-tail
coordinate (by convention here, $\xi=0$ corresponds to the head of the witness
bunch, and thus this bunch extends in the region $\xi < 0$),
$n_b(\xi)$ is the bunch density at position $\xi$, and $n_p$ is the plasma
density. Physically, the left-hand side of Eq.~(\ref{eq:evolution}) is the
equation of motion in a purely cylindrically symmetric wakefield, while the
right-hand side captures the asymmetric perturbations to the wakefield, which are driven by
the transverse offset $x_c$ of the beam (e.g. \cite{SchroederPRE2012,HuangPRL2007}).
The expression of the coefficients $k_\beta(\xi)$, $k_c$ and $\kappa_p$
depend on the wakefield regime considered.

For instance, in the blow-out regime \cite{HuangPRL2007},
Eq.~(\ref{eq:evolution}) applies with
$k_\beta^2(\xi) = k_p^2/2\gamma$, $k_c^2 = (n_{p}/n_{b,0}) k_p^2/2\gamma$
and $\kappa_p = \sqrt{n_{b,0} r_b^2/n_{p} r_0^2}\; (c_\psi k_p/\sqrt{2})$
(assuming a uniform bunch density $n_b(\xi)=n_{b,0}$),
where $k_p$ is the plasma wavevector, $r_b$ and $\gamma$ are the
radius and Lorentz factor of the
monoenergetic witness beam, $r_0$ is the radius of the blown-out cavity,
and the coefficient $c_\psi$
(given in \cite{HuangPRL2007}) takes into account the relativistic nature of the electron sheath.
In this regime and for a monoenergetic bunch,
there are no head-to-tail variations of the betatron frequency, i.e. $k_\beta$
is independent of $\xi$.

On the other hand, in the quasi-linear regime (and for a witness
beam having a transverse flat-top profile with $k_p r_b \ll 1$),
Eq.~(\ref{eq:evolution}) applies with $\kappa_p = k_p$, $k_c^2 = k_p^2/2\gamma$
and
\begin{eqnarray}
\label{eq:kB}
k_\beta^2(\xi) = &&\frac{k_p^2}{2\gamma} \eta_{d\perp} \sin[ k_p(\xi_d-\xi) ] \nonumber \\
&&+\frac{k_p^2}{2\gamma}\int_{\xi}^0\! \frac{n_b(\xi')}{n_p} \sin[k_p(\xi'-\xi)] \,k_p d\xi'
\end{eqnarray}
where the first term corresponds to the transverse wakefield generated by the
driver (either a laser or a charged particle bunch) and the second term corresponds to
transverse beam loading by the witness beam \cite{KatsouleasPartAcc1987}.
In the above expression, $\xi_d$ is the average longitudinal position of the
driver and $\eta_{d\perp}$ is the amplitude of the transverse driver wakefield.
For example, in the case of a flat-top electron bunch driver with a density
$n_d$, radius $r_d$, and length $\ell_d$, this amplitude is
$\eta_{d\perp} = (n_d/n_p) \,k_p r_d K_1( k_p r_d ) \,2\sin(k_p\ell_d/2)$, where $K_1$
is the modified Bessel function. In case of a linearly-polarized
Gaussian laser pulse with an amplitude $a_0 \lesssim 1$,
a waist $w_0 \gg r_b$, and an RMS duration $\tau$, $\eta_{d\perp} =
\sqrt{8\pi}/(k_p w_0)^2 \times a_0^2 (\omega_p\tau) e^{-(\omega_p\tau)^2/2}$.
From Eq.~(\ref{eq:kB}), it is clear that $k_\beta(\xi)$ exhibits head-to-tail
variations, even for a monoenergetic beam. As shown below, these variations
can lead to a saturation of the hosing instability.

\emph{Analytical solution for a linear chirp -}
In the general case, the system Eqs.~(\ref{eq:evolution})-(\ref{eq:kB})
can only be solved numerically. However, in order to gain insight into the
saturation mechanism, we first study Eq.~(\ref{eq:evolution})
analytically, in the simplified case of a linear betatron head-to-tail chirp
and uniform beam density:
\begin{equation}
\label{eq:linear-chirp}
k_\beta(\xi) = k_{\beta,0} + (\dkb) \,\xi\;, \qquad n_b(\xi) = n_{b,0}
\end{equation}
where $\dkb$ is constant and quantifies the betatron chirp.
Note that, although we will eventually apply this analytical solution to the
case of a quasi-linear wakefield, we keep the generic
notations $k_c$ and $\kappa_p$ from Eq.~(\ref{eq:evolution}) here,
so that the analysis can be applied to other similar situations (e.g.
blow-out regime with linear energy chirp).

For small betatron chirps ($|(\partial_\xi k_\beta) \xi| \ll k_{\beta,0}$),
we find that standard Laplace transform and steepest descents techniques can
be applied to find $x_c(\xi,z)$.
Starting for instance from a uniform initial offset $x_c(\xi, z=0) = \delta x$,
asymptotic solutions can be found for the early stage and late stage of the hosing instability,
where the transition between these two asymptotic solutions
is determined by a characteristic length
\begin{equation}
\label{eq:Lsat}
\Lsat(\xi) = \left(\frac{n_{b,0}}{n_p}\frac{k_c^2 \kappa_p^2}{k_{\beta,0}|\dkb|^3 |\xi|}\right)^{1/2}.
\end{equation}

(See the Supplementary Material for a derivation of these solutions.)
For $z \ll \Lsat(\xi)$ (early stage), the asymptotic solution is given by
\begin{eqnarray}
\label{eq:sol-growth}
&&x_c(\xi, z) = \delta x\, \frac{
\cos\left[ k_\beta(\xi) z - \frac{3}{4}N(\xi, z) + \frac{\pi}{12}
\right] }{(6\pi)^{1/2}N(\xi,z)^{1/2}}\,e^{\frac{3\sqrt{3}}{4}N(\xi, z)} \nonumber \\
&&\mathrm{with}\quad N(\xi, z) = \left( \frac{n_{b,0}}{n_p} \frac{k_c^2 \kappa_p^2
              z|\xi|^2}{k_{\beta,0}} \right)^{1/3}
\end{eqnarray}
and corresponds to the well-known scaling of the hosing instability, whereby the amplitude of
the betatron oscillations grows exponentially with $z$.

On the other hand, for $z \gg \Lsat(\xi)$ (late stage), the form of the
solution depends on the sign of $\dkb$. For $\dkb>0$
(increasing betatron frequency from tail to head), the solution is
given by
\begin{eqnarray}
\label{eq:sol-saturated}
&& x_c(\xi, z) = \delta x\frac{\cos[\,k_\beta(\xi) z -
   \varphi(z)\,]}{(8\pi^2)^{1/4}\Nsat(\xi)^{1/2}}\,e^{\sqrt{2}\Nsat(\xi)} \\
&&\mathrm{with}\quad \left\{\begin{array}{l}
\Nsat(\xi) = N(\xi, \Lsat) = \left( \frac{n_{b,0}}{n_p} \frac{k_c^2 \kappa_p^2 |\xi|}{k_{\beta,0}\,\dkb}\right)^{1/2}\\
\varphi(z) = \frac{n_{b,0}}{n_p}\frac{k_c^2 \kappa_p^2}{2 (\dkb)^2 k_{\beta,0} z}
\end{array}\right. \nonumber
\end{eqnarray}
 and corresponds to betatron oscillations
that have a constant, saturated amplitude. For $\dkb<0$ (decreasing frequency from tail to head),
the solution is given by
\begin{eqnarray}
\label{eq:sol-decreasing}
x_c(\xi, z) &=& -\delta x\,
\frac{\sin( k_{\beta,0} z - \varphi(z))}{(32\pi^2)^{1/4}\Nsat(\xi)^{-1/2}}\,\frac{e^{\sqrt{2}\Nsat(\xi)}}{|\dkb\, z \,\xi|} \\
&& +\,\delta x \,\frac{\cos[\,k_\beta(\xi) z -
  \varphi(z)\,]}{(\pi^2/2)^{1/4}\Nsat(\xi)^{1/2}}
\cos\left(\sqrt{2}\Nsat(\xi)-\frac{\pi}{4}\right) \nonumber
\end{eqnarray}
and corresponds to betatron
oscillations that initially decrease as $1/z$ [first term in
Eq.~(\ref{eq:sol-decreasing})] and eventually saturate at a low
level [second term in Eq.~(\ref{eq:sol-decreasing})].

We note that \cite{CherninPartAcc1989,StupakovSLAC1997} considered
beam-breakup in conventional accelerators, and that similar
analytical solutions were found in the case of a
wakefield function that is linear in $\xi$ (i.e. where the $\sin$ function in
Eq.~(\ref{eq:evolution}) is replaced by a linear function.) However this
short-beam linear wakefield function does not properly describe plasma
accelerators where, typically, $k_p\xi \sim 1$.

\begin{figure}
\includegraphics[width=\columnwidth]{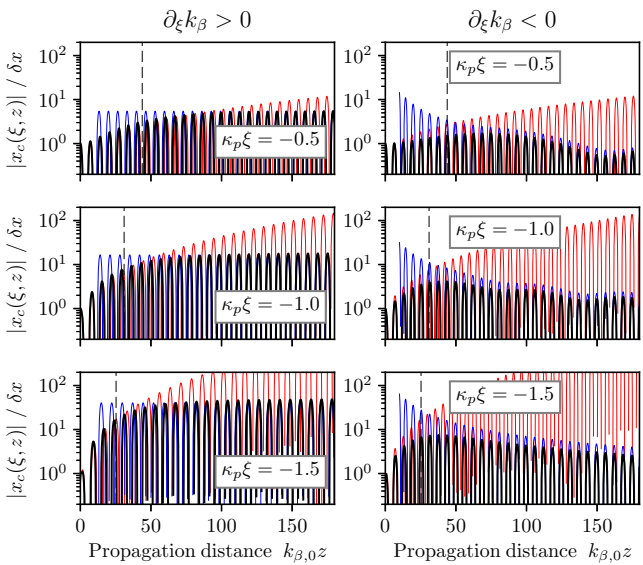}
\caption{\label{fig:comparison-analytical}
Comparison of the numerical integration
of the hosing equation Eq.~(\ref{eq:evolution}) in the case of a linear chirp
(black curve) with the early-stage solution (Eq.~(\ref{eq:sol-growth}); red curve) and
the late-stage solution (Eq.~(\ref{eq:sol-saturated}) and
Eq.~(\ref{eq:sol-decreasing}) on the left and right panels respectively;
blue curve), at different slices along the witness bunch (i.e. different
values of $\kappa_p\xi$). The vertical dashed line corresponds to $z=\Lsat(\xi)$,
from Eq.~(\ref{eq:Lsat}). The parameters used here are $k_c^2 n_{b,0}/n_p = k_{\beta,0}^2$
and $\dkb=0.1\,k_{\beta,0}\kappa_p$ (left panels) or $\dkb=-0.1\,k_{\beta,0}\kappa_p$ (right panels).}
\end{figure}

The asymptotic solutions Eqs.~(\ref{eq:sol-growth}), (\ref{eq:sol-saturated})
and (\ref{eq:sol-decreasing}) are compared with the explicit numerical
integration of the hosing equation Eq.~(\ref{eq:evolution}) in
Fig.~\ref{fig:comparison-analytical}, for $\partial_\xi k_{\beta}>0$ (left panels)
and $\partial_\xi k_{\beta}<0$ (right panels). As expected, in both cases
the numerical solution (black curve) is initially in agreement
with the standard scaling Eq.~(\ref{eq:sol-growth}) (red curve).
For longer propagation distances, the instability saturates and is in good agreement
with Eq.~(\ref{eq:sol-saturated}) and Eq.~(\ref{eq:sol-decreasing}), respectively (blue curves).
Additionally, Eq.~(\ref{eq:Lsat}) correctly predicts the approximate position of
the transition between the early-stage and late-stage regimes (vertical dashed lines).

Thus, positive and negative chirps ($\dkb>0$ and $\dkb<0$) exhibit qualitatively different
behaviors, but both strongly mitigate the hosing instability (compared to the case with no chirp).
In these cases, using the standard scaling
Eq.~(\ref{eq:sol-growth}) -- which does not take into account this mitigation --
can lead to an overestimation of the instability by an order of magnitude, or more.
Qualitatively, this is because a betatron chirp causes the different slices of
the bunch to progressively dephase and disrupts their coherent contribution
to the instability after a length $\Lsat$.
Note that the scaling  $\Lsat \propto |\dkb|^{-3/2}$
cannot be obtained from a coarse two-particle model \cite{ChaoHandbook}.

We note that the qualitative behavior for $\dkb<0$, whereby
the amplitude of the oscillations initially increases but later decreases,
is consistent with the behavior shown in \cite{MehrlingPRL2017},
where the authors observed, in numerical simulations, that
the deceleration of the tail of
driver bunch (which induces $\dkb<0$) in the blow-out regime could cause
a similar decrease in amplitude.

\emph{Quasi-linear wakefield with optimal beam loading -} Let us now connect the
simplified case of a linear betatron chirp (Eq.~(\ref{eq:linear-chirp}))
with the more generic case of the quasi-linear wakefield (Eq.~(\ref{eq:kB})).
Here we will consider the important
case where $n_b(\xi)$ is constrained by optimal beam loading.
In order to produce monoenergetic beams, it is indeed desirable to
tailor the beam density so as to flatten the accelerating field.
The accelerating force on a narrow witness electron beam
($k_p r_b \ll 1$) is given by \cite{KatsouleasPartAcc1987}
\begin{eqnarray}
\label{eq:Fz}
&&\frac{F_z(\xi)}{m c \omega_p} =  -\eta_{d\parallel}\cos[ k_p(\xi_d - \xi) ]  \\
&&- [1 - k_p r_b K_1(k_p r_b)]\int_\xi^0 \!\!
\frac{n_b(\xi')}{n_p} \cos(k_p(\xi'-\xi))\,k_pd\xi' \nonumber
\end{eqnarray}
where $\eta_{d\parallel}$ is the amplitude of the longitudinal driven wakefield.
For example, $\eta_{d\parallel} = [1 - k_p r_d K_1(k_p r_d)] (n_d/n_p)2\sin(k_p\ell_d/2)$
for a flat-top electron bunch driver, and
$\eta_{d\parallel} = \sqrt{\pi/8}\times a_0^2 (\omega_p\tau) e^{-(\omega_p\tau)^2/2}$
for a Gaussian laser pulse. In these conditions, it is well-known \cite{KatsouleasPartAcc1987}
that optimal beam-loading (i.e. uniform accelerating field) is obtained for a triangular-shaped witness bunch,
with (in our notation)
\begin{equation}
\label{eq:nb_beamloading}
n_b(\xi) = \frac{n_p\eta_{d\parallel}}{[1 - k_p r_b K_1(k_p r_b)]}[\,\sin(k_p\xi_d) - \cos(k_p\xi_d)k_p\xi\,]
\end{equation}
Inserting this expression into the equations for the accelerating
force (Eq.~(\ref{eq:Fz})) and the betatron frequency (Eq.~(\ref{eq:kB})) yields
\begin{eqnarray}
&&F_z(\xi) = - \eta_{d\parallel} mc\omega_p \cos(k_p\xi_d) \\
&&k_\beta^2(\xi) = \frac{k_p^2}{2\gamma}
\frac{\eta_{d\parallel}}{[1 - k_p r_b K_1(k_p r_b)]}
[ \, \sin(k_p\xi_d) - \cos(k_p\xi_d)k_p\xi \,] \nonumber \\
&& \;\;\;+ \frac{k_p^2}{2\gamma}\eta_{d\perp}\left(1  -
\frac{(\eta_{d\parallel}/\eta_{d\perp})}{[1 - k_p r_b K_1(k_p r_b)]}\right) \sin(k_p(\xi_d-\xi)) \label{eq:kB-optimal}
\end{eqnarray}

According to Eq.~(\ref{eq:kB-optimal}), in the case of optimal
beam loading, the form of the head-to-tail variations of
$k_\beta(\xi)$ depends on the ratio of the transverse and longitudinal
driven wakefield $\eta_{d\parallel}/\eta_{d\perp}$, and thus on
the shape of the driver. For instance, for $\eta_{d\parallel}/\eta_{d\perp} =
1 - k_p r_b K_1(k_p r_b)$
 (which occurs e.g. for a narrow bunch driver with a radius $r_d$ equal to that
 of the witness beam $r_b$), the second term in Eq.~(\ref{eq:kB-optimal})
 vanishes, and so $k_\beta(\xi)^2$ is simply linear in $\xi$, with a slope proportional
 to $-\cos(k_p\xi_d)$. Note that, in the accelerating phase of the wakefield,
 $\cos(k_d \xi_d)<0$, and thus this situation corresponds to the regime
 $\partial_\xi k_{\beta} > 0$. On the other hand, for
 $\eta_{d\parallel}/\eta_{d\perp} \gg 1 - k_p r_b K_1(k_p r_b)$
 (which is usually the case for a Gaussian laser driver, or a wide bunch driver) the variations
 of $k_{\beta}(\xi)$ are more complicated, with $\partial_\xi
 k_{\beta}$ changing sign from head to tail.

In order to illustrate these two situations, we carried out Particle-In-Cell (PIC) simulations
with (a) a bunch driver having a radius $r_d = 1.25 \,r_b$ (which corresponds to
 $\eta_{d\parallel}/\eta_{d\perp} \approx 1 - k_p r_b K_1(k_p r_b)$,
 and thus $\dkb > 0$) and (b) a
 laser driver having $k_p w = 2$ (which corresponds to
  $\eta_{d\parallel}/\eta_{d\perp} \gg 1 - k_p r_b K_1(k_p r_b)$, and
  thus $\dkb<0$ in most of the bunch).
For simplicity, we disabled driver evolution in the simulations, and
imposed a driver velocity $v_d = c$. (This is valid for acceleration distances
shorter than the characteristic lengthscale of driver evolution,
which is on the order of $\gamma_d r_d^2/\epsilon_d$ for a beam driver with
a normalized emittance $\epsilon_d$, and on the order of the dephasing
length $k_0^2/k_p^3$ for a guided laser driver.)
In both simulations (which used $n_p = 2\times 10^{17}\,\mathrm{cm}^{-3}$), a witness beam
with $\gamma = 200$ was initialized with a longitudinal density $n_b(\xi)$ given by
Eq.~(\ref{eq:nb_beamloading}) (i.e. optimal beam loading), and a transverse Kapchinskij-Vladimirskij
distribution \cite{LundPRSTAB2009} with a radius $r_b = 3 \,\mathrm{\mu m}$ and which was matched
to $k_\beta(\xi)$ as given by Eq.~(\ref{eq:kB-optimal}) (thereby ensuring that the transverse density profile
remains close to flat-top throughout the simulation). In order to seed the hosing instability, the witness bunch
was shifted transversally by an initial uniform offset $x_c(\xi,0)= 0.12\,\mathrm{\mu m}$.
The simulations were performed with the spectral quasi-cylindrical code FBPIC \cite{LeheCPC2016}
using the azimuthal modes $m=0$ and $m=1$. (This is sufficient because the
fields of the unperturbed, symmetrical beam are entirely contained in the mode $m=0$, and
because the perturbations due to the small offset $x_c$ create additional contributions
in the modes $m>0$ with a typical amplitude $(x_c/r_b)^m$ \cite{SchroederPRE2012}. Thus,
if $x_c$ is small compared to $r_b$, the relevant physics can be captured by the leading order, i.e. the mode $m=1$.)
In the simulations, the cell size was $\Delta z = 0.17 \;\mathrm{\mu m}$ and
$\Delta r = 0.06 \;\mathrm{\mu m}$, the timestep was chosen such that
$c\Delta t = \Delta z$, and the background plasma was represented with
8 macroparticles per cell (which was sufficient to reach numerical
convergence, as ascertained by separate tests featuring 32
macroparticles per cell).

\begin{figure}
\includegraphics[width=\columnwidth]{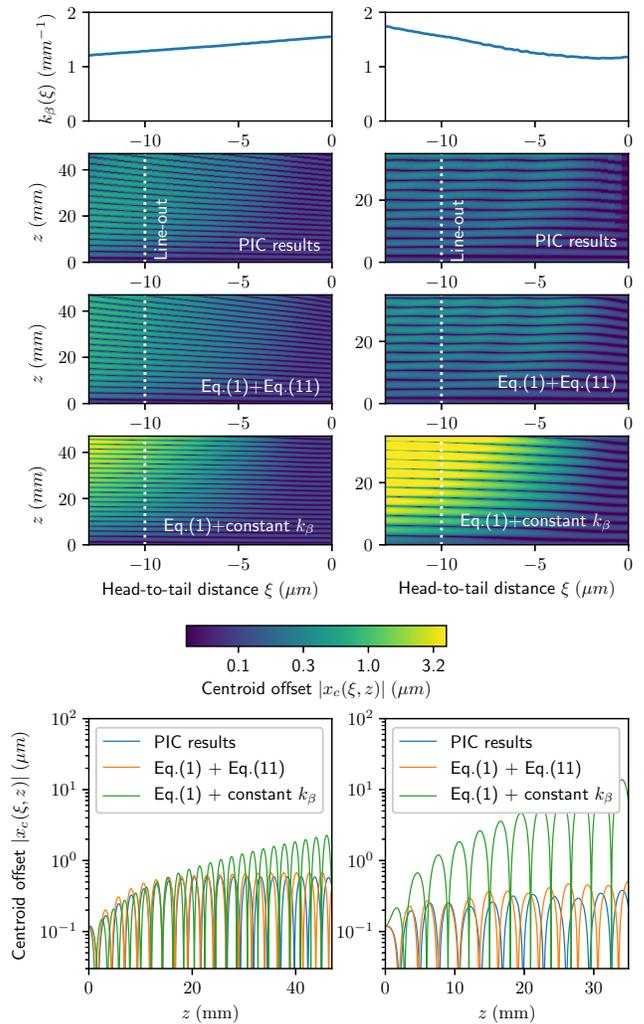}
\caption{\label{fig:PIC}
Evolution of the centroid offset (log-scaled colormap) as a function of the head-to-tail coordinate
$\xi$ and the propagation distance $z$, for (case (a); left panels) a bunch driver
with $r_d=4\,\mu m$, $\ell_d=3\,\mu m$, $\xi_d = 27\,\mu m$ and $n_d=0.7 \,n_p$, and (case (b); right panels)
a laser driver with $a_0 = 0.4$, $\tau = 20\,\mathrm{fs}$, $\xi_d = 26 \,\mu m$, $w = 24 \, \mu m$.
The top plots show the head-to-tail variations of the betatron frequency $k_\beta$ in both cases.
The upper colormaps show results from the PIC simulations, while the middle and lower colormaps
show results from the numerical integration of Eq.~(\ref{eq:evolution}), with $k_\beta(\xi)$
either given by Eq.~(\ref{eq:kB-optimal}) (middle colormaps) or considered uniform (lower colormaps).
The lower plots are line-outs of the colormaps, at a fixed head-to-tail distance
($\xi = -10\;\mathrm{\mu m}$; indicated by a white dotted line on the colormaps).
}
\end{figure}

The PIC simulation results are shown in the upper panels of Fig.~\ref{fig:PIC}. As
expected from the analysis of Eq.~(\ref{eq:kB-optimal}), the betatron frequency
$k_\beta$ exhibits head-to-tail variations in both cases (top plots).
This causes the instability to quickly saturate as a
function of $z$ in the PIC simulations (upper colormaps, showing that the
maximum centroid offset reaches only a limited value on the logarithmic
colorscale), in a
way that is consistent with the numerical integration of the equation of hosing
Eq.~(\ref{eq:evolution}) with $k_\beta$ given by Eq.~(\ref{eq:kB-optimal})
(middle colormaps). Importantly,
the level of the instability is much lower than it would have been in the case
of a uniform $k_\beta$ (lower colormaps, reaching higher values on the logarithmic
colorscale). This can also be seen on the line-outs of the colormaps at a fixed position
$\xi$ (bottom plots), which, again, show that the PIC simulations and
Eq.~(\ref{eq:evolution}) with Eq.~(\ref{eq:kB-optimal}) are
in agreement regarding the amplitude of the centroid oscillations (some
differences occur because Eq.~(\ref{eq:evolution}) neglects beam acceleration
and the evolution of the transverse beam profile), and that this amplitude is
lower than that predicted by Eq.~(\ref{eq:evolution}) with a constant $k_\beta$.
This confirms that, in the quasi-linear regime, the hosing instability
is less severe than suggested by the standard scalings (e.g. Eq.~(\ref{eq:sol-growth}))
that assume a uniform $k_\beta$.

In conclusion, we showed that, in the quasi-linear regime of plasma acceleration,
the hosing instability is strongly mitigated, even for a monoenergetic bunch.
This is due to the natural variations of the focusing forces across the bunch,
and happens both for increasing and decreasing head-to-tail variations.
In the case of optimal beam loading, the exact form of these head-to-tail
variations is controlled by the shape of the driver.

\begin{acknowledgments}
This work was supported by the Director, Office of Science, Office of High Energy Physics,
of the U.S. Department of Energy under Contract
No. DE-AC0205CH11231. Simulations
were performed on the Lawrencium computational cluster resource
provided by the IT Division at the Lawrence Berkeley National
Laboratory (Supported by the Director, Office of Science, Office of
Basic Energy Sciences, of the U.S. Department of Energy under Contract
No. DE-AC02-05CH11231).
\end{acknowledgments}

%

\end{document}